\documentclass{article}
\usepackage{ijcai26}

\usepackage{times}
\usepackage{soul}
\usepackage{url}
\usepackage[hidelinks]{hyperref}
\usepackage[utf8]{inputenc}
\usepackage[small]{caption}
\usepackage{graphicx}
\usepackage{amsmath}
\usepackage{amssymb}
\usepackage{amsthm}
\usepackage{booktabs}
\usepackage{multirow}
\usepackage{algorithm}
\usepackage{algorithmic}

\usepackage{subcaption}
\usepackage{booktabs}
\usepackage{tcolorbox}
\usepackage{fontawesome5} 
\tcbuselibrary{skins, breakable}

\title{TrajAD: Trajectory Anomaly Detection for Trustworthy LLM Agents}

\author{
Yibing Liu$^1$
\and
Chong Zhang$^1$\and
Zhongyi Han$^1$\thanks{*Corresponding author: Zhongyi Han, Email: hanzhongyicn@gmail.com}\and
Hansong Liu$^2$\and
Yong Wang$^2$\and
Yang Yu$^3$\and
Xiaoyan Wang$^4$\And
Yilong Yin$^1$\thanks{*Corresponding author: Yilong Yin, Email: ylyin@sdu.edu.cn}
\affiliations
$^1$School of Software, Shandong University, Jinan, China\\
$^2$Sonli Holding Group Co., Ltd., Qingdao, China\\
$^3$Shandong Huazhi Talent Technology Co., Ltd., Jinan, China\\
$^4$Information Technology Service Center of People's Court\\
\emails
sduliuyb@163.com,
zhangchongupc@163.com,
hanzhongyicn@gmail.com,
liuhansong@sonli.net,
wangyong@sonli.net,
yuy@sdas.org,
428163395@139.com,
ylyin@sdu.edu.cn
}

\begin{document}

\maketitle

\begin{abstract}
We address the problem of runtime trajectory anomaly detection, a critical capability for enabling trustworthy LLM agents. Current safety measures predominantly focus on static input/output filtering. However, we argue that ensuring LLM agents reliability requires auditing the intermediate execution process. In this work, we formulate the task of Trajectory Anomaly Detection. The goal is not merely detection, but precise error localization. This capability is essential for enabling efficient rollback-and-retry. To achieve this, we construct TrajBench, a dataset synthesized via a perturb-and-complete strategy to cover diverse procedural anomalies. Using this benchmark, we investigate the capability of models in process supervision. We observe that general-purpose LLMs, even with zero-shot prompting, struggle to identify and localize these anomalies. This reveals that generalized capabilities do not automatically translate to process reliability. To address this, we propose TrajAD, a specialized verifier trained with fine-grained process supervision. Our approach outperforms baselines, demonstrating that specialized supervision is essential for building trustworthy agents.
\end{abstract}

\begin{figure*}[t]
        \centering
        \includegraphics[width=1\linewidth]{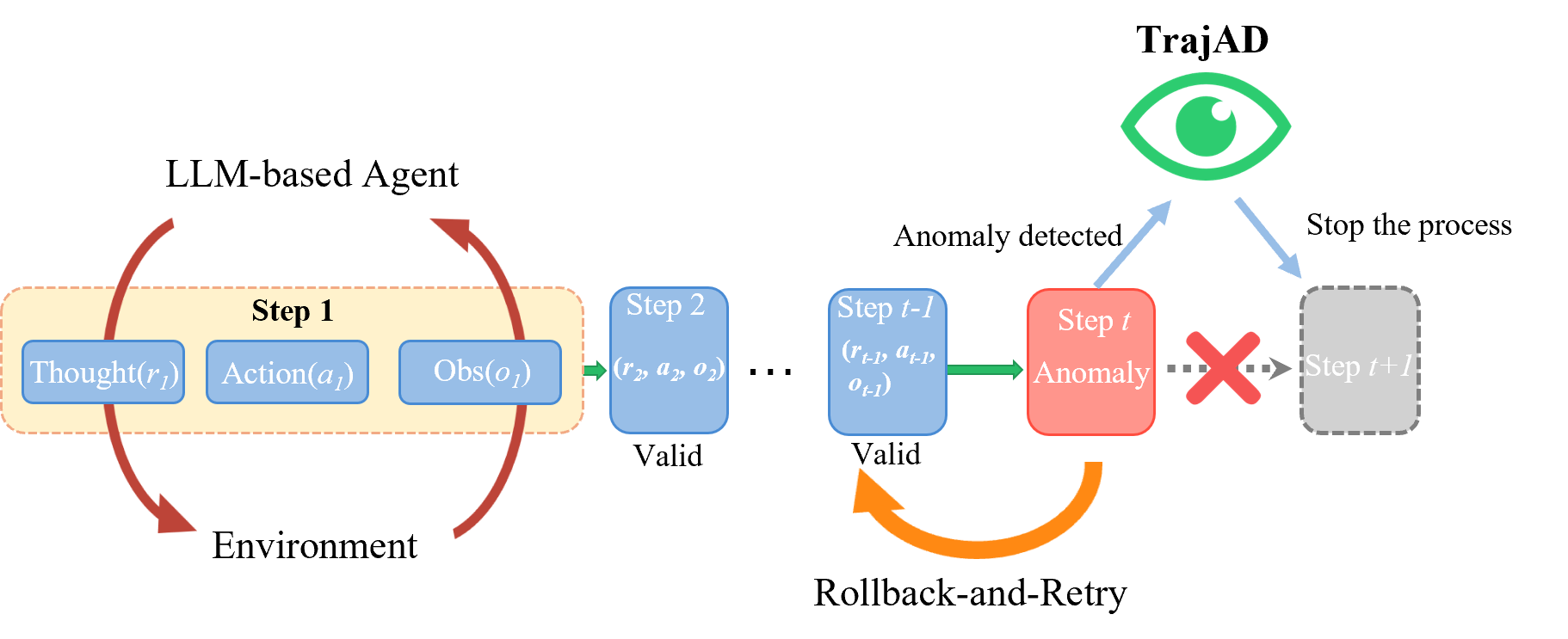}
        \caption{Overview of the TrajAD framework. 
        We introduce the TrajAD framework to verify the agent's trajectories. At each step, the agent generates a thought, takes an action, and receives an observation, forming an execution unit. The execution trajectory is periodically validated to check whether it remains normal. If all previous steps are valid, execution continues. When an anomaly is detected at step \textit{t}, the process is halted before step \textit{t+1}. The trajectory can rollback to the step \textit{t-1} and retry instead of restarting the whole task.}
        \label{fig:framework}
\end{figure*}

\section{Introduction}

LLM-based agents function as autonomous systems that leverage reasoning and planning to decompose complex goals into executable steps~\cite{park2023generativeagent}. They have demonstrated potential in high-stakes domains, such as financial decision-making~\cite{wang2023fingpt,zhou2024finrobot} and clinical diagnosis~\cite{singhal2023large,tang2024medagents}, where precision is paramount. However, despite this progress, the widespread deployment is hindered by safety concerns. In these safety-critical environments, the lack of robustness in the execution process poses severe risks.

A critical challenge is the risk of trajectory anomalies. An agent's execution involves complex interleaving of reasoning, tool usage, and environmental feedback. Due to this complexity, agents often commit errors in intermediate steps. Common anomalies include fabricating invalid tool parameters, entering infinite loops, or executing redundant actions that are locally plausible but globally inefficient. Crucially, these anomalies do not always result in immediate task failure. However, they lead to significant resource waste and potential safety risks, such as irreversible database corruption~\cite{yuan2024rjudge,ying2025securewebarena}. This necessitates a mechanism to detect anomalies in the execution process and interrupt errors in real-time.

Current efforts primarily focus on enhancing capabilities or static safety, neither of which effectively addresses runtime trajectory anomalies. On the capability side, methods like trajectory-based fine-tuning~\cite{chen2023fireact,zeng2024agenttuning,song2024agentbank} and Process Reward Models (PRM)~\cite{lightman2023PRM} introduce supervision during the training phase. However, they aim to optimize model parameters to improve the general policy. They do not function as a runtime monitor to audit specific execution instances. Similarly, safety measures such as hallucination detection~\cite{zhang2025siren,huang2025survey} and safety guardrails~\cite{zhang2024agent,dong2025safeguarding} operate on a local or static scope. They typically verify the final output against facts or filter atomic tool calls in isolation. Crucially, these methods lack temporal awareness. They fail to detect logical errors inherent to the sequence, such as infinite loops or redundant actions. Moreover, they cannot localize the specific error step. This necessitates a dedicated mechanism to verify the entire execution trajectory.

However, achieving this goal faces two primary obstacles. First, there is a lack of datasets that contrast normal trajectories with anomalous ones. Current datasets~\cite{zeng2024agenttuning,song2024agentbank} rely on gold-standard trajectories to provide positive supervision. They rarely include annotated ``negative samples" which are essential for learning to identify anomalies. Just as humans learn from mistakes, models require exposure to failure modes to establish robust decision boundaries. Second, precise anomaly detection and localization present significant challenges. Existing methods are optimized for instruction following and task completion. However, they are not explicitly designed to differentiate between normal and anomalous behaviors. Furthermore, accurately pinpointing the exact position of an error is difficult. The boundary between a complex reasoning step and a redundant loop is often ambiguous without deep semantic understanding. Solving this localization problem significantly improves efficiency. Agents can ``rollback" to the error step instead of restarting the entire task.

To address these challenges, we propose a systematic framework designed to proactively verify the execution process (Figure~\ref{fig:framework}). We formally define the task of Trajectory Anomaly Detection. This task requires the model to distinguish anomalies based on global trajectory context. Crucially, it must also localize the exact error step to enable subsequent recovery. To achieve this, we construct TrajBench. We employ ``Perturb-and-Complete" strategies on gold-standard trajectories to generate high-quality negative samples. We combine these with normal trajectories to form TrajBench. This dataset enables models to distinguish anomalies and locate errors. Using this dataset, we fine-tune Qwen3-4B to develop TrajAD as our core detector. Experiments show that general-purpose models struggle to distinguish anomalous trajectories. They fail even more on the challenging task of localizing exact error steps. In contrast, TrajAD achieves superior performance in both detection and localization.

Our main contributions are as follows:
\begin{itemize}
    \item We identify and formalize the problem of Agent Trajectory Anomaly Detection. To the best of our knowledge, this is the first work to systematically investigate procedural anomalies in agent execution. We shift the evaluation paradigm from outcome-centric correctness to process-centric rationality, highlighting the critical need for runtime auditing mechanisms.
    \item We construct TrajBench, the first high-quality dataset dedicated to agent execution anomalies. It covers three representative anomaly categories: Task Failure, Process Inefficiency, and Unwarranted Continuation. To ensure generalization, we build upon AgentBank, which covers five core dimensions: reasoning, mathematics, coding, web navigation, and embodied AI. We modify these expert trajectories to synthesize corresponding fine-grained anomalies.
    \item We propose TrajAD, a specialized auditing framework for detecting and localizing anomalies. By modeling the global context of execution traces, TrajAD achieves precise step-level localization of errors. This capability enables efficient error recovery through a ``rollback-and-retry" mechanism, significantly improving agent reliability and reducing resource consumption.
\end{itemize}

\section{Related Work}

\subsection{Advancements in LLM-based Agents}

Recent advancements in agentic systems primarily follow two paradigms based on architectural design and parameter update. Architectural design enhances LLM-based agents without modifying model weights. Reasoning frameworks decompose complex tasks into sequential thought processes~\cite{wei2022cot,yao2023tot}. ReAct~\cite{yao2022react} synergizes reasoning with acting by interleaving thoughts with observations. Memory mechanisms retrieve external knowledge~\cite{lewis2020RAG} and past experiences~\cite{zhang2025ACE} to extend long-term memory. Furthermore, the Model Context Protocol (MCP)\footnote{\url{https://modelcontextprotocol.io}} standardizes tool integration, allowing agents to plug into dynamic environments directly. Conversely, the parameter update paradigm embeds capabilities directly into the LLM's inherent knowledge. General approaches align models with broad user intents via instruction tuning~\cite{DBLP:conf/iclr/WeiBZGYLDDL22} and knowledge distillation~\cite{hinton2015distilling}. Going further, Process Reward Models (PRMs)~\cite{lightman2023PRM} introduce step-level supervision using human-labeled intermediate states. Trajectory-based fine-tuning ~\cite{chen2023fireact,zeng2024agenttuning,song2024agentbank} optimize models on interaction trajectories. This enables agents to master the skills required for autonomous tasks.

However, these advancements prioritize capability over reliability. While agents can now handle harder problems, they exhibit ``blind goal-directedness"~\cite{shayegani2025BGD}. They greedily optimize for final outcomes while neglecting process rationality. This renders the execution process uncontrollable and risky. With increased autonomy, trajectory anomalies pose severe risks. Structural loops unnecessarily deplete computational budgets. More critically, unverified actions can trigger irreversible state changes, such as corrupting databases or executing unauthorized transactions. In safety-critical domains, such unstable behaviors undermine trust, making the agent unsafe regardless of the final outcome. Current methods focus on task success but lack the ability to verify their own intermediate steps. 

\subsection{Trustworthiness in Agents}

Current research on agent trustworthiness primarily centers on hallucination detection, safety guardrails, and LLM-as-a-Judge. Hallucination detection targets factual correctness, checking textual consistency~\cite{manakul2023selfcheckgpt} or execution validity~\cite{chern2023factool} against ground truth. Safety guardrails deploy external filters~\cite{inan2023llamaguard} or programmable rules~\cite{rebedea2023nemo} to intercept adversarial attacks, preventing both toxic content generation, malicious prompts, and risky tool invocations~\cite{yuan2024rjudge,ying2025securewebarena}. The ``LLM-as-a-Judge" paradigm utilizes strong generalist models to grade the quality of generated content against human preferences~\cite{zheng2023LLM-as-a-Judge,bai2022constitutionalAI,liu2023geval}.

However, these methods fail to monitor the dynamic execution process. Hallucination detection relies on static textual checks, while safety guardrails serve as passive defenses against external attacks. Similarly, current LLM judges rely on prompting for zero-shot evaluation. They lack the domain-specific knowledge to identify subtle anomalies within the process. In contrast, we synthesize anomaly trajectories to fine-tune a specialized verifier. This enables the model to grasp the logical dependencies between steps, allowing it to accurately identify anomalies and locate the exact step.

\section{Problem Formulation}
\label{sec:problem_formulation}

In this section, we establish the formal framework for Trajectory Anomaly Detection. We first model the agent execution as a sequential decision process. Unlike outcome-centric evaluations, we characterize anomalies based on \textit{process rationality}, focusing on three primary categories of anomalies. Finally, we formulate the auditing task as a supervised learning problem, where the objective is to jointly predict the anomaly verdict and localize the specific error step.

\subsection{Preliminaries}
We formulate agent execution as a sequential decision process. Given a task instruction $I$, the agent interacts with the environment over $n$ steps.  At each step $t$, the agent first generates a thought $r_t$ for planning. Conditioned on this reasoning, it executes an action $a_t$. Upon receiving the action, the environment transitions to a new state and returns an observation $o_t$ reflecting this change. This interaction cycle repeats until the task is completed. We define the interaction trajectory $T$ as the sequence of these triplets:
\begin{equation}
  T = \{I,(r_1, a_1, o_1),(r_2, a_2, o_2), \dots,(r_n, a_n, o_n)\},   
\end{equation}
where $n$ denotes the total number of steps. This formulation explicitly captures the interleaving of reasoning, execution, and feedback.

\subsection{Taxonomy of Anomalies}
\label{taxonomy}
Unlike outcome-based evaluation, we focus on the rationality of the execution process. We focus on three primary categories of anomalies:
\begin{itemize}
    \item \textbf{Type I: Task Failure ($\mathcal{A}_{fail}$).}
        The agent fails to complete the task. This includes two cases:
        \begin{enumerate}
            \item[(a)] \textit{Reasoning Error}: The agent executes a valid action $a_t$ based on flawed reasoning $r_t$.
            \item[(b)] \textit{Execution Error}: The agent executes an incorrect action $a_t$, causing runtime exceptions.
        \end{enumerate}
    \item \textbf{Type II: Process Inefficiency ($\mathcal{A}_{ineff}$).} The agent completes the task, but with redundant steps. Formally, a trajectory is inefficient if a shorter trajectory $T'$ exists that achieves the same outcome (i.e., $|T'| < |T|$). This includes circular loops or extra actions that are locally plausible but globally inefficient.
    \item \textbf{Type III: Unwarranted Continuation ($\mathcal{A}_{unw}$).} The agent fails to stop when tasks are impossible or unnecessary due to environment changes.
    \begin{enumerate}
        \item[(a)] \textit{Failure to Refuse}: The task is impossible under current constraints. The agent fails to report the inability and hallucinates a plan.
        \item[(b)] \textit{Redundant Continuation}: The task is already finished or the context has changed, making further actions meaningless. The agent fails to perceive this termination condition and continues execution.
    \end{enumerate}
\end{itemize}

\subsection{Task Definition}
We define Trajectory Anomaly Detection as a supervised auditing task. Given a trajectory $T$, the goal is to learn a mapping function $f: T \to (c, l)$.

Here, $c \in \{\text{Normal}, \text{Anomaly}\}$ represents the binary verdict of the trajectory's validity. The variable $l$ denotes the \textit{First Error Step}:
\begin{equation}
    l = 
    \begin{cases} 
    t_{err}, & \text{if } c = \text{Anomaly}; \\
    \emptyset, & \text{if } c = \text{Normal},
    \end{cases}
\end{equation}
where $t_{err} \in \{1, \dots, n\}$ is the index of the first step where the anomaly occurs. Precise prediction of $l$ is critical, as it enables the agent to rollback to the pre-error state $s_{l-1}$ for efficient recovery, rather than restarting the entire task.

\begin{figure*}
    \centering
    \includegraphics[width=1\linewidth]{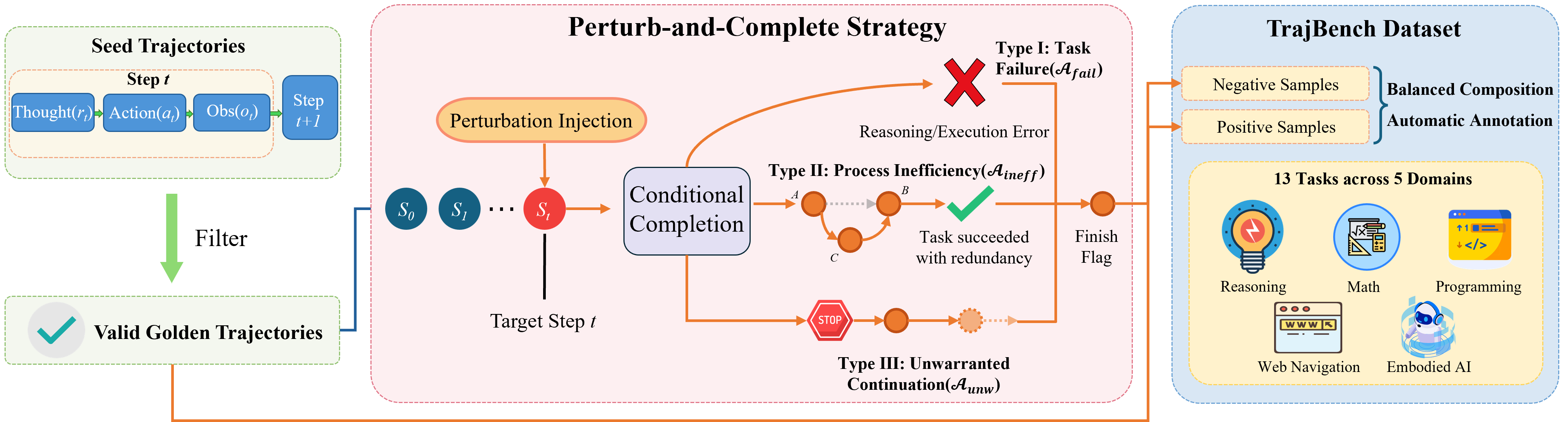}
    \caption{The data construction pipeline for TrajBench.
    We initialize the process by filtering seed trajectories to ensure a high-quality set of valid golden trajectories. To construct negative samples, we employ a Perturb-and-Complete strategy. We inject a perturbation into a target step $S_t$ and force conditional completion to finish the subsequent trajectory based on the altered context. This process synthesizes three distinct anomaly types: Task Failure ($\mathcal{A}_{fail}$), Process Inefficiency ($\mathcal{A}_{ineff}$), and Unwarranted Continuation ($\mathcal{A}_{unw}$). The resulting dataset features a balanced composition of positive and negative samples, where anomalous trajectories are automatically annotated. The dataset consists of 13 tasks across 5 domains.}
    \label{fig:dataset}
\end{figure*}

\section{TrajBench: A Dataset for Trajectory Anomaly Detection}
\label{sec:trajbench}

To enable the auditing task defined in Sec.~\ref{sec:problem_formulation}, a dataset containing both execution anomalies and precise error localization is required. Existing benchmarks~\cite{song2024agentbank} primarily target imitation learning, consisting solely of expert demonstrations. They lack the negative samples and step-wise annotations necessary for learning process verification. To bridge this gap, we construct TrajBench, a large-scale dataset synthesized via a semi-automated pipeline. TrajBench explicitly pairs golden trajectories with strictly defined anomalies, providing full supervision for both the anomaly verdict $c$ and error location $l$.

\subsection{Data Construction Pipeline}

We utilize AgentBank~\cite{song2024agentbank} as our seed dataset due to its broad coverage across five core domains: Reasoning, Math, Programming, Web Navigation, and Embodied AI. To ensure the quality of the base data, we first employ a validator model to filter the raw trajectories, retaining only those with logically sound reasoning chains. Based on these verified seeds, we apply a Perturb-and-Complete strategy to synthesize negative samples. This process involves three steps:

\paragraph{Step 1: Perturbation Injection.}
Given a golden trajectory $T_{gold}$, we sample a target step $t$. To ensure the verifier captures global context, we prioritize sampling from intermediate positions rather than early steps. We inject a perturbation into step $t$ to create a deviated state, strictly following the anomaly taxonomy in Sec.~\ref{taxonomy}:

\begin{itemize}
    \item \textbf{Type I: Task Failure ($\mathcal{A}_{fail}$).} We inject fatal errors into the execution stream. For \textit{Reasoning Errors}, we replace the valid thought $r_t$ with a logical flaw or state misconception. For \textit{Execution Errors}, we modify action $a_t$ to invoke incorrect tools or invalid parameters.
    
    \item \textbf{Type II: Process Inefficiency ($\mathcal{A}_{ineff}$).} We introduce redundancy without altering the final outcome. We insert semantically valid but useless sub-sequences (e.g., loops $A \to B \to A$ or detours $A \to C \to B$) into the trajectory. These actions appear locally plausible but waste computational resources, challenging the verifier to identify global inefficiency.
    
    \item \textbf{Type III: Unwarranted Continuation ($\mathcal{A}_{unw}$).} We manipulate the termination conditions. To simulate \textit{Failure to Refuse}, we remove necessary tools or set conflicting constraints, forcing the agent to hallucinate a plan. To simulate \textit{Redundant Continuation}, we inject a ``Task Completed'' signal into observation $o_t$ but instruct the agent to ignore it and continue execution.
\end{itemize}

\paragraph{Step 2: Conditional Completion.}
Conditioned on the perturbed history $T_{\le t}$, a strong language model generator is employed to simulate the subsequent behavior $T_{>t}$. We explicitly constrain the generation to maintain logical consistency with the injected error (e.g., continuing a wrong path after a reasoning error) to ensure the trajectory remains coherent.

\paragraph{Step 3: Automatic Annotation.}
A key advantage of this pipeline is the acquisition of precise labels without manual cost. Since the perturbation step $t$ is controlled, we automatically assign the ground truth error location $L_{loc} = t$ and the verdict $C_{verdict} = \text{Anomaly}$. The valid seed trajectory serves as the positive sample ($C_{verdict} = \text{Normal}$).

\subsection{Dataset Statistics and Quality Analysis}
\label{sec:dataset_analysis}

TrajBench comprises a total of 60,000+ trajectories, strictly balanced with a 1:1 ratio between normal and anomalous samples. The dataset covers 13 tasks across five domains, ensuring broad diversity. The anomalies are evenly distributed, with Type I, II, and III accounting for roughly 33\% each (see Section 1 of the Supplementary Material for details).

To ensure high quality, we implement a rigorous two-stage verification process. During the seed collection phase, we employed a validator model to verify the logical consistency of the source trajectories. Only samples with coherent reasoning chains were retained, resulting in a pass rate of 91.6\% (retaining 34436 out of 37625 raw seeds). This ensures that our positive samples are strictly ``golden". From these verified seeds, our pipeline successfully synthesized 31,742 valid anomalous trajectories (a generation success rate of 92.2\%). The remaining failures were due to model refusal or format parsing errors. To establish a rigorous evaluation setting, we construct TrajBench by pairing each successfully synthesized anomaly with its original source trajectory. This results in a strictly balanced dataset of 63,484 samples, eliminating class distribution bias.

Finally, to assess the reliability of the synthesized labels, we conduct a human review on a stratified random subset of 500 samples (100 per domain). Annotators verify two criteria: (1) anomaly category alignment, and (2) error step localization precision. The results show a Human-Model Agreement rate of 96.2\% for classification and 94.5\% for localization. This high consistency confirms that our automated pipeline produces trusted supervision signals.

\section{TrajAD: A Generative Verifier for Agent Trajectories}

We propose TrajAD, a generative verifier designed for the auditing task defined in Sec.~\ref{sec:problem_formulation}. We formulate the problem as conditional text generation and fine-tune the model on the TrajBench dataset (Sec.~\ref{sec:trajbench}). Formally, the input sequence $\mathcal{X}$ consists of a system instruction $I_{sys}$ and the trajectory $T$. Given an input sequence $\mathcal{X} = \{I_{sys}, T\}$, the model generates a structured diagnostic report $\mathcal{Y}$:
\begin{equation}
\mathcal{Y} = [C_{cls}; L_{loc}], 
\end{equation}
where $C_{cls} \in \{\text{Normal}, \text{Anomaly}\}$ denotes the verdict and $L_{loc} \in \{1, \dots, n\}$ denotes the index of the error step.

We adopt a standard decoder-only Transformer as the backbone. To adapt the model to the auditing task while maintaining computational efficiency, we employ Low-Rank Adaptation (LoRA)~\cite{hu2022lora}. We freeze the pre-trained weights $W_0$ and introduce trainable low-rank matrices $A$ and $B$. The forward pass is modulated as $h = (W_0 + BA)x$. We optimize the parameters $\Phi = \{A, B\}$ using the standard autoregressive objective over the output $\mathcal{Y}$ in TrajBench:
\begin{equation}
\mathcal{L} = -\sum_{t=1}^{|\mathcal{Y}|} \log P(y_t | \mathcal{X}, y_{<t}), 
\end{equation}
This formulation allows the model to learn the joint distribution of anomaly verdicts and error locations directly from the supervision signals provided in TrajBench.

During inference, TrajAD operates as a runtime monitor embedded in the agent's execution loop (Figure~\ref{fig:framework}). We perform verification at a fixed step interval. The model takes the current trajectory history as input and predicts the tuple $(C_{cls}, L_{loc})$. The inference process follows a ``Check-and-Act" protocol: If $C_{cls} = \text{Normal}$, the agent continues execution. If $C_{cls} = \text{Anomaly}$, the execution is interrupted. The agent then utilizes the predicted index in $L_{loc}$ to rollback the environment to the pre-error state $s_{l-1}$, enabling targeted recovery without a full restart.

\section{Experiments}
\label{sec:experiments}

We evaluate TrajAD on the TrajBench dataset to validate its effectiveness in verifying agent trajectories. Our experiments focus on three key questions: (1) Does specialized trajectory detection outperform general-purpose reasoning? (2) Can the model robustly localize errors across diverse domains? (3) How does data scale impact the verification capability?

\begin{table*}[t]
\centering
\caption{Main Results: The overall performance comparison of TrajAD against baseline models. We report Precision ($\mathcal{P}$), Recall ($\mathcal{R}$), and Macro-F1 ($\mathcal{F}_1$) for anomaly detection, and Joint Exact Match (JEM) for error step localization. The best results are highlighted in \textbf{bold}, and the second-best results are \underline{underlined}.}
\label{tab:main_results}
\resizebox{\textwidth}{!}{%
\begin{tabular}{l|c|c|ccc|c}
\toprule
\multirow{2}{*}{\textbf{Model}} & \multirow{2}{*}{\textbf{Params}} & \multirow{2}{*}{\textbf{Method}} & \multicolumn{3}{c|}{\textbf{Anomaly Detection(\%)}} & \textbf{Localization(\%)} \\ \cmidrule(lr){4-6} \cmidrule(lr){7-7} 
 &  &  & \textbf{Precision}($\mathcal{P}$) & \textbf{Recall}($\mathcal{R}$) & \textbf{Macro-F1}($\mathcal{F}_1$) & \textbf{Joint Exact Match(JEM)} \\  \midrule
\multicolumn{7}{l}{\textit{\textbf{Open-Source Baselines}}} \\ 
Gemma-3-4B-Instruct & 4B & Zero-shot & 68.64 & 64.66 & 64.20 & \underline{9.07} \\ 
Phi-3-Mini & 4B & Zero-shot & 67.78 & 28.46 & 30.65 & 3.28 \\ 
Qwen3-4B & 4B & Zero-shot & \underline{79.07} & 68.97 & \underline{70.43} & 5.54 \\ 
Qwen3-8B & 8B & Zero-shot & 76.16 & \underline{69.60} & 67.90 & 5.81 \\ \midrule
\multicolumn{7}{l}{\textit{\textbf{Ours}}} \\ 
\textbf{TrajAD(Ours)} & 4B & LoRA finetune & \textbf{82.90} & \textbf{82.49} & \textbf{81.81} & \textbf{53.75} \\ \bottomrule
\end{tabular}%
}
\end{table*}

\subsection{Experimental Setup}

\paragraph{Dataset and Baselines.}
We utilize the balanced TrajBench dataset (60k samples) constructed in Sec.~\ref{sec:trajbench}. We adopt a stratified split, reserving 10\% of samples from each task for testing. We benchmark TrajAD (fine-tuned Qwen3-4B) against representative zero-shot baselines. We select models to evaluate distinct hypotheses: \begin{itemize} 
    \item \textbf{Qwen3-4B (Base)~\cite{yang2025qwen3} \& Gemma-3-4B-Instruct~\cite{kamath2025gemma}:} As general-purpose models of the same scale, they serve to evaluate whether standard instruction-following capabilities are sufficient for anomaly auditing without specialized supervision. 
    \item \textbf{Phi-3-Mini-4k-Instruct~\cite{abdin2024phi3technicalreporthighly}:} We include this lightweight model to benchmark the reasoning capabilities of small-scale LLMs across the full dataset. 
    \item \textbf{Qwen3-8B~\cite{yang2025qwen3}:} We include a larger-scale model to investigate whether simply scaling model capacity can solve the auditing challenge without specific fine-tuning.
\end{itemize}

\paragraph{Evaluation Metrics.}
We employ a multi-dimensional evaluation suite (see the Supplementary Material for complete definitions and details):
\begin{itemize}
    \item \textbf{Detection (Binary Classification):} We report Precision ($\mathcal{P}$), Recall ($\mathcal{R}$), and Macro-F1 ($\mathcal{F}_1$). We prioritize Recall to minimize safety risks associated with missed anomalies.
    \item \textbf{Localization (Joint Verification):} We define a strict \textbf{Joint Exact Match (JEM)} metric. A prediction is considered correct if and only if:
    \begin{enumerate}
        \item The predicted error step index $l_{pred}$ exactly matches the ground truth $l_{gt}$.
        \item The semantic similarity between the generated error content $c_{pred}$ and the ground truth $c_{gt}$ exceeds a threshold $\tau$.
    \end{enumerate}
    Formally, $\text{JEM} = \mathbb{I}(l_{pred} = l_{gt}) \cdot \mathbb{I}(\text{sim}(c_{pred}, c_{gt}) > \tau)$. We compute similarity using the Ratcliff-Obershelp algorithm (via Python's \texttt{difflib} module). We set $\tau=0.2$ to allow for diverse phrasing while rejecting irrelevant content. Our preliminary verification experiments reveal that models can achieve artificially high localization scores (e.g., $\sim 71.5\%$). A closer inspection of the outputs indicates that models often guess the correct index without identifying the actual anomaly location. JEM evaluates whether the model correctly identifies \textit{why} a step is anomalous, rather than simply guessing \textit{where} it is.
\end{itemize}

\paragraph{Implementation Details.}
We fine-tune the Qwen3-4B base model using QLoRA~\cite{dettmers2023qlora}. We attach Low-Rank Adapters ($r=8, \alpha=16$) to all linear layers, affecting only 1.8\% of total parameters. Training employs the Paged AdamW 8-bit optimizer with a peak learning rate of $2\times 10^{-5}$ and a 10\% warmup. All experiments are conducted on a single NVIDIA A100 (80GB) GPU.

\begin{figure}[t]
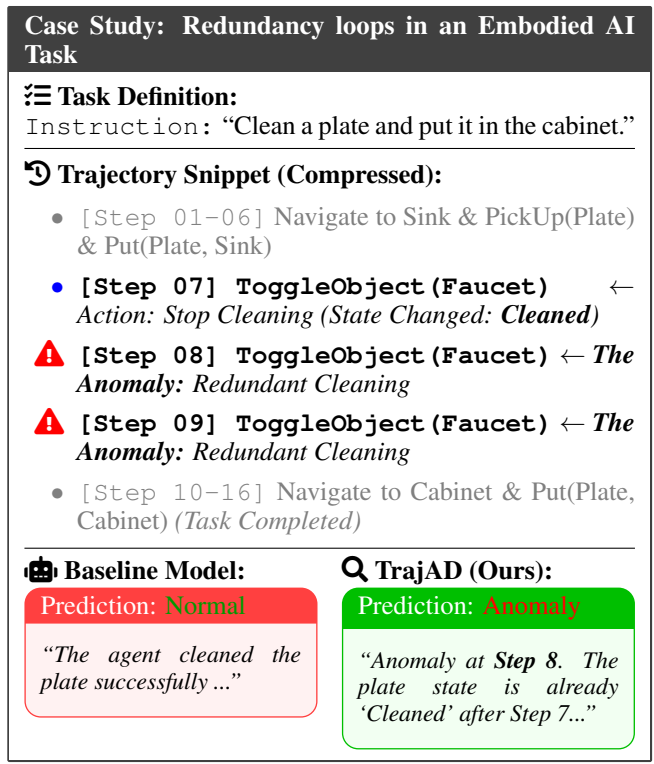

    \centering
    \tcbset{
        colback=white, colframe=black!75!white, 
        fonttitle=\bfseries, arc=0mm, boxrule=0.8pt,
        left=1mm, right=1mm, top=0.4mm, bottom=0.4mm
    }
    
    \begin{tcolorbox}[title=Case Study: Redundancy loops in an Embodied AI Task]
        
        \textbf{\faTasks\ Task Definition:} \\
        \texttt{Instruction:} ``Clean a plate and put it in the cabinet." \\
        \vspace{-0.2cm}
        \hrule
        \vspace{0.15cm}
        
        \textbf{\faHistory\ Trajectory Snippet (Compressed):}
        
        \begin{itemize}
            \item[\textcolor{gray}{$\bullet$}] \textcolor{gray}{\texttt{[Step 01-06]} Navigate to Sink \& PickUp(Plate) \& Put(Plate, Sink)}
            \item[\textcolor{blue}{$\bullet$}] \textbf{\texttt{[Step 07] ToggleObject(Faucet)}} \hfill \textit{$\leftarrow$ Action: Stop Cleaning (State Changed: \textbf{Cleaned})}
            
            \item[\textcolor{red}{\faExclamationTriangle}] \textbf{\texttt{[Step 08] ToggleObject(Faucet)}} \hfill \textit{$\leftarrow$ \textbf{The Anomaly:} Redundant Cleaning}
            \item[\textcolor{red}{\faExclamationTriangle}] \textbf{\texttt{[Step 09] ToggleObject(Faucet)}} \hfill \textit{$\leftarrow$ \textbf{The Anomaly:} Redundant Cleaning}
            \item[\textcolor{gray}{$\bullet$}] \textcolor{gray}{\texttt{[Step 10-16]} Navigate to Cabinet \& Put(Plate, Cabinet) \textit{(Task Completed)}}
        \end{itemize}
        \hrule
        \vspace{0.1cm}
        
        \begin{minipage}[t]{0.48\linewidth}
            \textbf{\faRobot\ Baseline Model:}
            \vspace{-2mm}
            
            \begin{tcolorbox}[colback=red!5!white, colframe=red!75!white, title=Prediction: \textcolor{green!60!black}{Normal}, arc=2mm, boxrule=0.5pt, left=1mm, right=1mm]
                \small
                \textit{``The agent cleaned the plate successfully ..."}
            \end{tcolorbox}
        \end{minipage}
        \hfill
        \begin{minipage}[t]{0.48\linewidth}
            \textbf{\faSearch\ TrajAD (Ours):}
            \vspace{-2mm}
            
            \begin{tcolorbox}[colback=green!5!white, colframe=green!75!black, title=Prediction: \textcolor{red}{Anomaly}, arc=2mm, boxrule=0.5pt, left=1mm, right=1mm]
                \small
                \textit{``Anomaly at \textbf{Step 8}. The plate state is already `Cleaned' after Step 7..."}
            \end{tcolorbox}
        \end{minipage}
        
    \end{tcolorbox}
    \vspace{-0.4cm}
    \caption{Qualitative Comparison on Redundancy Loops. The baseline model overlooks the repeated cleaning action since it does not affect the final goal state. In contrast, TrajAD detects the redundancy as a Inefficiency process.}
    \label{fig:case}
\end{figure}

\subsection{Main Results}
\label{sec:in_distribution}

We first evaluate the model's performance under the In-Distribution (ID) setting, where the training and testing samples are drawn from the same set of 13 tasks across 5 domains. Note that while the tasks are seen, the specific test trajectories are strictly held-out.

Table~\ref{tab:main_results} summarizes the performance. TrajAD achieves a substantial improvement over all baselines, validating the effectiveness of trajectory anomaly detection. As shown in Table~\ref{tab:main_results}, zero-shot models exhibit a critical Precision-Recall imbalance. For instance, Qwen3-4B achieves a high Precision of 79.07\% but a low Recall of 68.97\%, while Phi-3 suffers from a severe Recall collapse (28.46\%). This indicates a conservative bias. Pre-trained models tend to assume agent actions are valid, failing to detect subtle anomalies. This leads to a high false-negative rate. Furthermore, their localization capability is virtually non-existent, with Joint Exact Match (JEM) scores consistently below 10\%. This indicates that general-purpose LLMs lack the capability to precisely localize errors within long trajectory sequences. As shown in Figure~\ref{fig:case}, baselines often overlook subtle procedural anomalies (e.g., redundant actions) as long as the final goal is achieved.

In contrast, TrajAD effectively overcomes these limitations. It improves Macro-F1 by 11.38\% (to 81.81\%) compared to the strongest baseline. More importantly, it achieves a breakthrough in localization, boosting JEM by 48.21\% (to 53.75\%). This demonstrates that our generative objective successfully forces the model to couple logical reasoning with structural verification, enabling precise error diagnosis.

\begin{figure}
    \centering
    \includegraphics[width=1\linewidth]{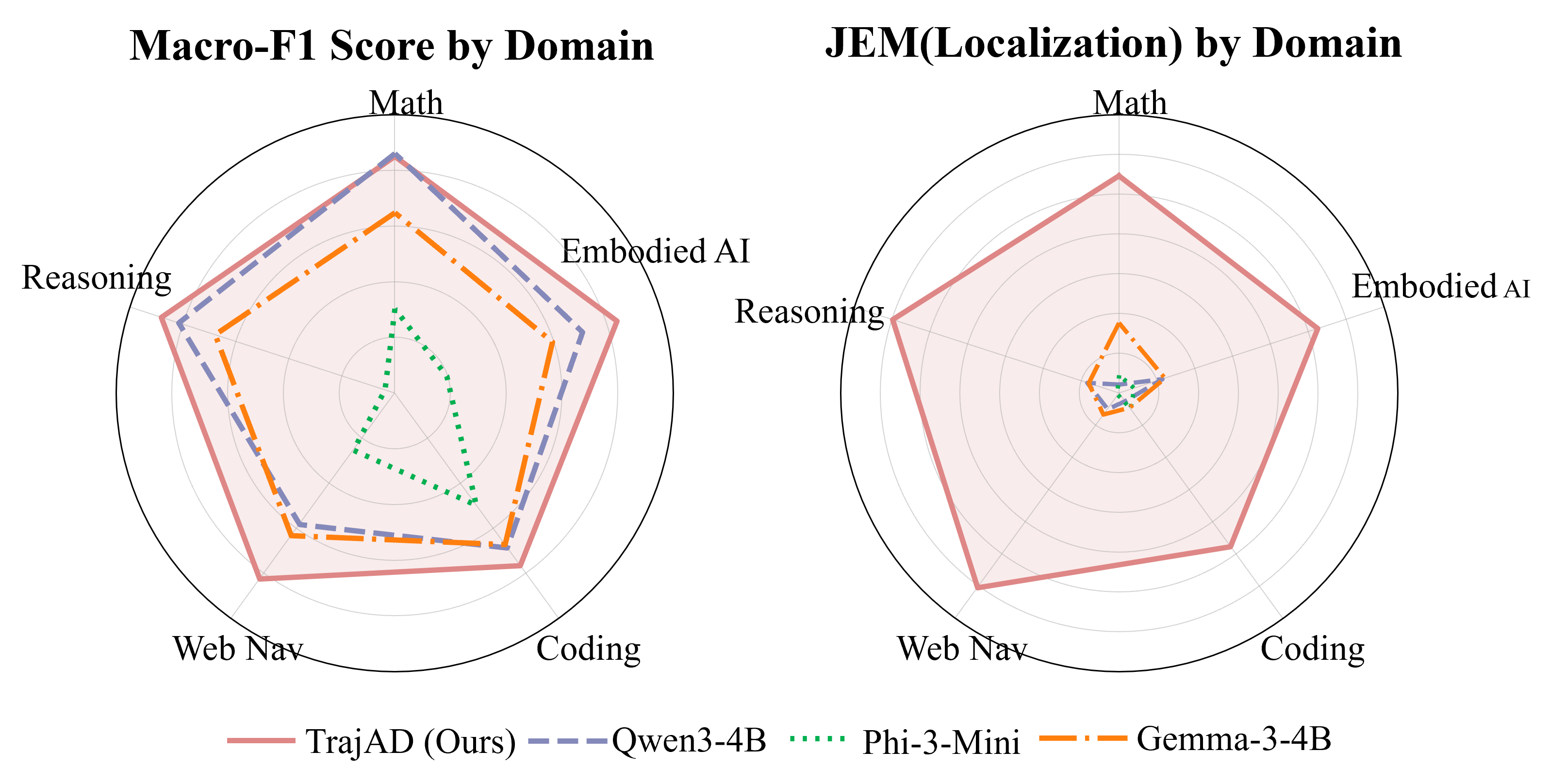}
    \vspace{0.1cm}
    \caption{Domain-Specific Performance Analysis. (Left) Macro-F1 across five domains. TrajAD (solid red) forms the outermost envelope, demonstrating consistent robustness. (Right) Exact Match. Baselines exhibit a structural collapse near the center, highlighting their inability to localize errors, whereas TrajAD maintains a functional verification boundary.}
    \label{fig:radar}
    \vspace{0.1cm}
\end{figure}

To analyze performance stability, we decompose the ID evaluation into five domains: Math, Reasoning, Coding, Web Navigation, and Embodied AI. Figure~\ref{fig:radar} illustrates the results on each domain. TrajAD consistently outperforms baselines across all five domains. Zero-shot baselines fail to ground actions in Embodied AI tasks, resulting in near-zero localization. In contrast, TrajAD maintains high precision in this complex domain. This confirms that under the ID setting, our method successfully masters the distinct verification logic required for each domain.

\begin{figure}[t]
  \centering
  \begin{subfigure}{\linewidth}
    \centering
    \includegraphics[width=\linewidth]{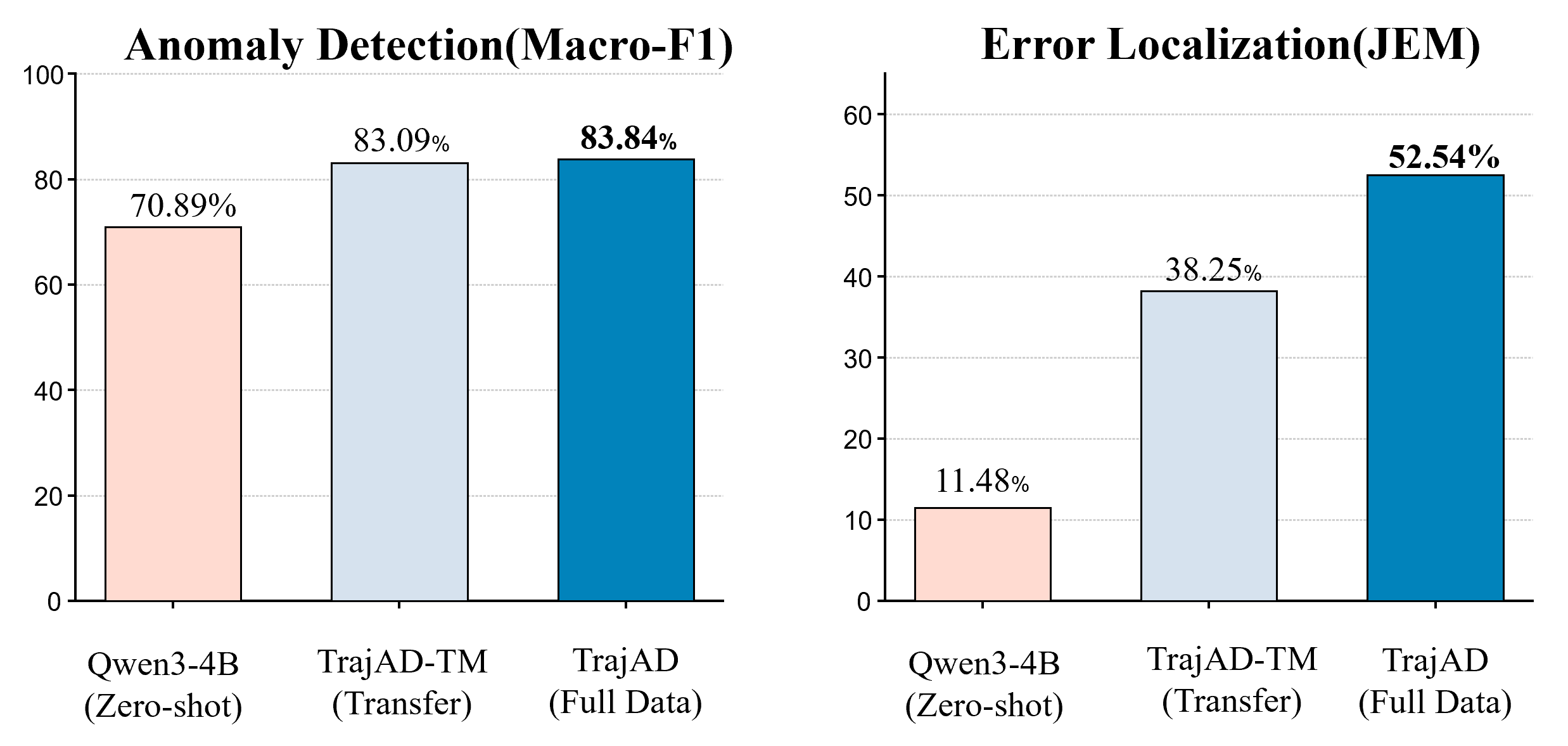} 
    \caption{Generalization Capabilities. TrajAD (Transfer) demonstrates strong zero-shot detection performance on the held-out Embodied AI domain, though localization benefits from in-domain training.}
    \label{fig:analysis_generalization}
  \end{subfigure}
  \hfill
  \begin{subfigure}{\linewidth}
    \centering
    \includegraphics[width=\linewidth]{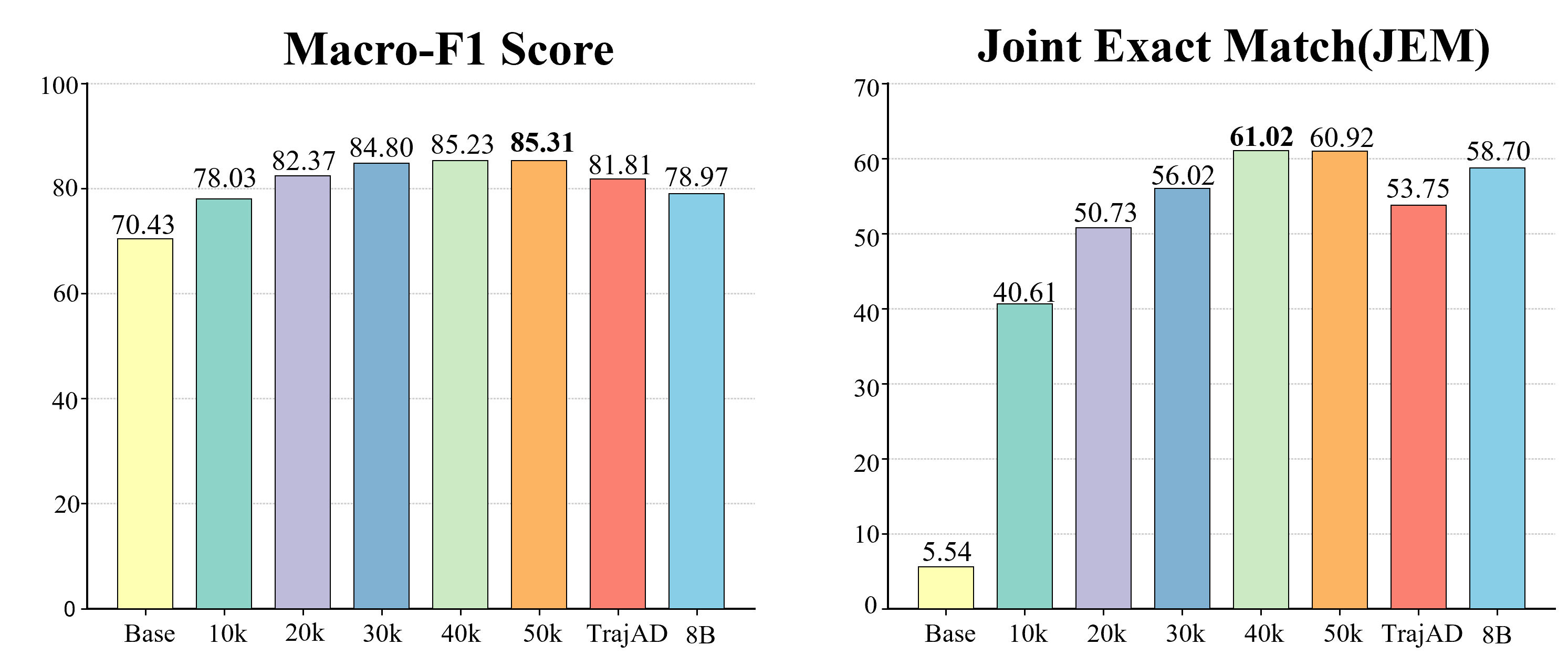}
    \caption{Scaling Law \& Model Capacity. Performance improves linearly with data scale up to 50k samples. Scaling to 60k introduces negative transfer, and increasing model size (8B) does not overcome this distribution bottleneck.}
    \label{fig:analysis_scaling}
  \end{subfigure}
  \vspace{-0.2cm}
  \caption{Ablation and Analysis Experiments. 
  (a) Generalization: Comparison of zero-shot transfer (Transfer Model) vs. full supervision. The detection gap is minimal, validating the universality of the learned logic.
  (b) Scalability: Impact of training data size and model parameters. The 4B model with 50k stratified samples achieves optimal efficiency, outperforming both the full-data 4B model and the larger 8B baseline.}
  \label{fig:comprehensive_analysis}
\end{figure}

\subsection{Out-of-Distribution Generalization}
\label{sec:ood_generalization}

A critical question is whether TrajAD learns universal verification logic or simply memorizes domain-specific patterns. To investigate this, we conduct a Cross-Domain Transfer experiment under the Out-of-Distribution (OOD) setting.

To ensure the target task is unseen, we adopt a strict Leave-One-Domain-Out protocol. We select Embodied AI as the held-out target domain $D_{target}$ and train a transfer model, TrajAD-TM, on the remaining source domains $D_{source} = \{\text{Math, Reasoning, Coding, Web}\}$. We then evaluate this model directly on $D_{target}$ without any further adaptation. This rigorous setting tests the model's ability to transfer verification logic to a novel action space without relying on memorized domain patterns.

As shown in Figure~\ref{fig:analysis_generalization}, TrajAD-TM exhibits strong transferability, outperforming the zero-shot baseline on the unseen domain. Specifically, it improves Macro-F1 from 70.89\% to 83.09\% and JEM from 11.48\% to 38.25\%. However, a performance gap remains when compared to the fully supervised upper bound. While detection performance is nearly identical (83.09\% vs. 83.84\% F1), localization still lags behind the supervised model (38.25\% vs. 52.54\% JEM). This indicates that localization is more sensitive to subtle variations in anomaly patterns across domains. This sensitivity suggests that TrajAD can serve as a probe to extract domain-specific failure modes, facilitating targeted improvements in agents.

\subsection{Scaling and Efficiency Analysis}
\label{sec:scaling}

Finally, we investigate the efficiency of our framework by analyzing the impact of training data scale and model capacity. We fine-tune TrajAD on stratified subsets ranging from 10k to 60k samples. As illustrated in Fig.~\ref{fig:analysis_scaling}, performance correlates positively with data size in the early stages. Increasing samples from 10k to 50k yields consistent improvement, peaking at 85.31\% F1 and 61.02\% JEM.

To investigate model capacity constraints, we extend our evaluation to the larger Qwen3-8B model. First, in the zero-shot setting (Table~\ref{tab:main_results}), the 8B base model fails to outperform the 4B base model (67.90\% vs. 70.43\% F1), suggesting that raw parameter count alone does not confer an advantage in auditing logic. We further fine-tune the 8B model on the full dataset. As shown in Fig.~\ref{fig:analysis_scaling}, this larger model achieves 78.97\% F1. Notably, this does not surpass the TrajAD model trained on the same data partition (81.81\% F1), nor does it reach the optimal 4B checkpoint (85.31\% F1). Even with fine-tuning, scaling up the model yields limited gains. This suggests that parameter scale is not the primary bottleneck for this task. Consequently, the 4B model demonstrates a superior trade-off between performance and computational cost, making it the more efficient choice for deployment.

\section{Conclusion}

In this work, we have addressed the challenge of ensuring agent reliability by formally defining the task of Trajectory Anomaly Detection. We construct TrajBench, the first large-scale benchmark dedicated to this purpose. Our experiments show that general-purpose LLMs fail to localize errors, regardless of model scale. They lack the capability to link detected anomalies to specific execution steps. We show that scaling up model size does not solve this issue. Specialized supervision is necessary. Our method, TrajAD, outperforms larger baselines. This proves that a small model is effective when trained to explicitly generate the error details. We hope this work serves as a foundational step, shifting the focus of agent evaluation from outcome-based metrics to rigorous process auditing, ultimately paving the way for trustworthy autonomous systems with minimal human intervention.

\clearpage

\bibliographystyle{named}
\bibliography{ijcai26}

\end{document}